\documentclass[11pt,sort&compress]{elsarticle}
\PassOptionsToPackage{sort&compress}{natbib}
\journal{}
\usepackage{lipsum}
\usepackage{amsfonts}
\usepackage{graphicx}
\usepackage{epstopdf}
\usepackage{algorithmic}
\usepackage{multirow}
\usepackage{algorithm}
\usepackage{algorithmic}

\usepackage{epsf}
\usepackage{amsmath}
\usepackage{amssymb}
\usepackage{setspace}
\usepackage{enumerate}
\usepackage{float}
\usepackage{footnote}
\usepackage{scalefnt}
\usepackage{xfrac}
\usepackage{transparent}
\usepackage{rotate}
\usepackage{array}
\usepackage{tabu}
\usepackage[mediumspace,mediumqspace,squaren]{SIunits}
\usepackage{multirow}
\usepackage{enumitem}
\usepackage{tikz}
\usepackage{bm}
\usepackage{microtype}
\usepackage[margin=1in]{geometry}
\usetikzlibrary{patterns.meta}
\usepackage{outlines}
\usepackage{breqn}
\usepackage{etoolbox}

\usepackage{bm}
\newcommand{\ve}[1]{\bm{#1}}

\newcommand{\bu}{\ve{u}}

\newcommand{\bp}{\ve{p}}

\newcommand{\bc}{\ve{c}}

\newcommand{\bw}{\ve{w}}

\newcommand{\bg}{\ve{g}}

\newcommand{\bA}{\ve{A}}

\newcommand{\dd}{\text{d}}

\newcommand{\eps}{\varepsilon}

\newcommand\Rey{\mbox{\text{Re}}}

\newcommand\Ca{\mbox{\text{Ca}}} 
\newcommand{\overbar}[1]{\mkern 1.5mu\overline{\mkern-1.5mu#1\mkern-1.5mu}\mkern 1.5mu}

\usepackage{hyperref}
\usepackage[]{xcolor}

\hypersetup{
  colorlinks=true,
}

\definecolor{lightblue}{rgb}{0.63, 0.74, 0.78}
\definecolor{seagreen}{rgb}{0.18, 0.42, 0.41}
\definecolor{orange}{rgb}{0.85, 0.55, 0.13}
\definecolor{silver}{rgb}{0.69, 0.67, 0.66}
\definecolor{rust}{rgb}{0.72, 0.26, 0.06}
\definecolor{purp}{RGB}{68, 14, 156}

\colorlet{lightsilver}{silver!30!white}
\colorlet{darkorange}{orange!75!black}
\colorlet{darksilver}{silver!65!black}
\colorlet{darklightblue}{lightblue!65!black}
\colorlet{darkrust}{rust!85!black}
\colorlet{darkpurp}{purp!85!black}

\usepackage[labelfont=bf]{caption}
\usepackage[parfill]{parskip}

\usepackage[nameinlink]{cleveref}
\crefname{equation}{}{}
\Crefname{ALC@unique}{Line}{Lines}

\makeatletter
\preto\maketitle{%
  \begingroup\lccode`~=`,
  \lowercase{\endgroup
  \let\saved@breqn@active@comma~
  \let~}\active@comma 
}
\appto\maketitle{%
  \begingroup\lccode`~=`,
  \lowercase{\endgroup
  \let~}\saved@breqn@active@comma 
}
\makeatother

\begin{document}

\hypersetup{
  linkcolor=darkrust,
  citecolor=seagreen,
  urlcolor=darkrust,
  pdfauthor=author,
}

\begin{frontmatter}

\title{{\Large\bfseries Fast integration method for averaging \\ polydisperse bubble population dynamics}}

\author{\vspace{-4ex}Spencer H.\ Bryngelson\corref{cor1}}
\ead{shb@gatech.edu}
\address{\vspace{-2ex}School of Computational Science \& Engineering, Georgia Institute of Technology, Atlanta, GA 30332, USA}
\address{\vspace{-3ex}George W.\ Woodruff School of Mechanical Engineering, Georgia Institute of Technology, Atlanta, GA 30332, USA}
\address{\vspace{-3ex}Daniel Guggenheim School of Aerospace Engineering, Georgia Institute of Technology, Atlanta, GA 30332, USA}

\date{}

\begin{abstract}
    Ensemble-averaged polydisperse bubbly flow models require statistical moments of the evolving bubble size distribution.
    Under step forcing, these moments reach statistical equilibrium in finite time.
    However, the transitional phase before equilibrium and cases with time-dependent forcing are required to predict flow in engineering applications.
    Computing these moments is expensive because the integrands are highly oscillatory, even when the bubble dynamics are linear.
    Ensemble-averaged models compute these moments at each grid point and time step, making cost reduction important for large-scale bubbly flow simulations.
    Traditional methods evaluate the integrals via traditional quadrature rules.
    This approach requires a large number of quadrature nodes in the equilibrium bubble size, each equipped with its own advection partial differential equation (PDE), resulting in significant computational expense.
    We formulate a Levin collocation method to reduce this cost.
    Given the differential equation associated with the integrand, or moment, the method approximates it by evaluating its derivative via polynomial collocation.
    The differential matrix and amplitude function are well-suited to numerical differentiation via collocation, and so the computation is comparatively cheap.
    For an example excited polydisperse bubble population, the first moment is computed with the presented method at $10^{-3}$ relative error with 100 times fewer quadrature nodes than the trapezoidal rule.
    The gap increases for smaller target relative errors: the Levin method requires $10^4$ times fewer points for a relative error of $10^{-8}$.
    The formulated method maintains constant cost as the integrands become more oscillatory with time, making it particularly attractive for long-time simulations.
    Mechanistically, the transient behavior of the moments is set by two effects: resonance detuning across bubble sizes, which causes phase mixing of oscillations, and viscous damping, which removes radial kinetic energy.
    The proposed formulation isolates the oscillations while keeping the remaining terms smooth, so accuracy does not deteriorate at late times.
\end{abstract}

\end{frontmatter}

\section{Introduction}

Bubbly flows occur in numerous engineering applications, including ship propulsion systems~\citep{cook28}, hydraulic machinery~\citep{arndt81}, spillway flows~\citep{falvey90}, and biomedical applications such as lithotripsy and histotripsy~\citep{cleveland00,pishchalnikov03,maxwell11}.
In these systems, polydisperse bubble populations result from nucleation, breakup, and shedding of larger vapor regions~\citep{prosperetti17,risso18,deike22}.
The ensemble of bubble size oscillations characterizes the associated collective bubble dynamics.
Individual bubble dynamics can even cause shock waves comparable in magnitude to those in the carrier flow~\citep{reisman98,brennen95,trummler19}.
Further, small numbers of bubbles or small void fractions can change large-scale pressure wave propagation~\citep{mettin03}, and polydispersity meaningfully influences the acoustics in bubbly liquids~\citep{ando11,pishchalnikov03}.
For practical use in ship and biomedical settings, the implementation requires consistent boundary conditions at inlets and walls, prescribing the driving pressure signal at the inflow or outflow, specifying the bubble size distribution, and ensuring the ensemble moments are compatible with the carrier-flow boundary conditions.

Computational modeling of these flows presents challenges due to the range of scales.
Individual bubble radii span from microns to millimeters~\citep{brennen95}, while bubble or vapor clouds and turbulent structures extend to meters or larger depending on the application~\citep{d83}.
The temporal scales are disparate: bubble natural frequencies and collapse times occur on timescales of microseconds, while flow observation times extend to seconds or more~\citep{brennen95}.
This scale separation makes direct numerical simulation of resolved flow dynamics computationally expensive or intractable for engineering applications.

\begin{figure}[htbp]
	\centering
	\includegraphics{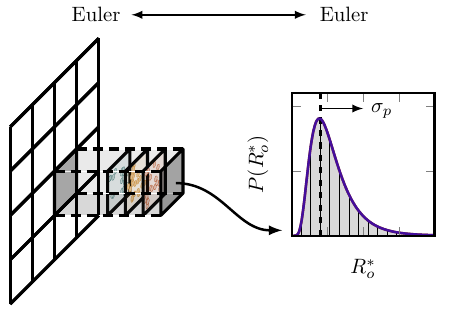}
	\caption{Schematic of ensemble-averaging approach for polydisperse bubble populations. Individual bubble dynamics are replaced by statistical moments of the bubble size distribution, which couple to the continuous liquid phase through mixture-averaged properties.}
	\label{f:classes}
\end{figure}

To address this challenge, ensemble-averaged sub-grid modeling approaches view the individual bubble dynamics as less important than the statistical properties of the bubble population~\citep{zhang94,zhang94b,ishii10,drew98}.
This view is referred to as two-way coupled Euler--Euler modeling and remains valid as long as the void fractions are sufficiently small.
Euler--Euler ensemble average models result in equations for the continuous liquid phase that are forced, in a two-way coupled manner, by statistical moments of bubble properties, such as radius and radial velocity.
The bubble variables become stochastic, Eulerian fields.
These methods contrast with Lagrangian approaches that track individual bubbles, or particles more broadly, as points with their own dynamics~\citep{bryngelson19,capecelatro13,crowe12}.
Recent advances in Eulerian multiphase closures~\citep{fox24,charalampopoulos21} develop well-posed models for polydisperse flows, while open-source frameworks~\citep{lehnigk22,wilfong252,bryngelson20_qbmm} provide computational tools for modeling the multiphase flow.
The ensemble-averaging approach is illustrated schematically in \cref{f:classes}, where the statistical properties of a bubble population determine the mixture-averaged flow properties.

The ensemble-averaged approach requires the computation of statistical moments of the bubble size distribution, which involve integrals over the bubble population.
Alternative approaches to handling polydispersity include quadrature-based moment methods (QMOM) and conditional quadrature methods (CQMOM)~\citep{marchisio13,fox08}, which evolve the moments directly through transport equations.
These methods are often used in the particle-laden flow community and, likewise, require closure assumptions for unclosed moments as well~\citep{fox08}.
The ensemble-averaged approach computes moments from the bubble dynamics.
Still, a computational challenge arises from the polydisperse nature of the bubble population: bubbles of different equilibrium sizes oscillate at different natural frequencies.
These frequencies are typically uncoupled from one another, and their distributions lead to complex oscillatory behavior in the bubble size distribution over time~\citep{ando11,ando10,bryngelson23}.
At a physical level, a polydisperse population behaves as an ensemble of damped radial oscillators.
The spread of natural frequencies across sizes drives dephasing and cancellation in the population-averaged moments, while viscosity sets the decay rate of these transients.
In the context of moment computation, this oscillatory behavior entails highly oscillatory integrands and, thus, expensive computation via traditional quadrature methods.

Conventional approaches to this problem employ large numbers of quadrature points to resolve the highly oscillatory integrands~\citep{bryngelson19_CPC}, or accept that the moment closure may not be accurate.
However, this strategy becomes computationally expensive; each additional quadrature point requires solving an additional advection partial differential equation (PDE).
For example, achieving acceptable accuracy can require thousands of quadrature points, as well as thousands of additional PDEs to accompany the Navier--Stokes-like PDEs.
These quadrature requirements mean the computational cost of moment evaluation exceeds that of the flow solver.

The challenge of computing integrals with very oscillatory integrands is not unique to bubbly flow modeling and has been recognized in the numerical analysis community.
Traditional quadrature methods, including Gaussian quadrature, exhibit poor convergence properties when the integrand oscillates more rapidly than the quadrature rule resolves, as they have not yet reached their asymptotic convergence rate.
\citet{filon28} first addressed this limitation by proposing trading numerical integration for differentiation.
\citet{levin96,levin97} extended the concept by developing an approach for constructing differential equations with non-oscillatory solutions that are solved efficiently using standard numerical methods.
Later developments include adaptive Filon methods~\citep{iserles05,iserles07}, stationary-phase approaches~\citep{olver08}, and analytic continuation techniques~\citep{huybrechs06}.
Reviews in this area suggest that specialized methods increase efficiency as the oscillation frequency increases~\citep{huybrechs17,sinha24neural}, making them attractive for problems such as polydisperse bubble dynamics, although many have constraints that make them inappropriate for the current work.
Quadrature-based moment methods (QMOM)~\citep{marchisio13,fox08} are widely used in the bubbly flow community for tracking dynamic processes.
Still, the problem addressed here involves a fundamentally different type of quadrature problem.
The independent variable $R_o$ represents a static distribution of equilibrium bubble sizes, not a dynamic variable that evolves in time, and thus is not amenable to QMOM-type approaches that track time-dependent variables.
While QMOM techniques have been extended to the internal coordinates (the bubble radii and their velocities) using traditional quadrature in $R_o$~\citep{bryngelson23}, the present work focuses on efficiently computing moments over the static equilibrium size distribution.

This work adapts the Levin collocation method to polydisperse bubble dynamics governed by Rayleigh--Plesset-type equations.
The key insight is that the oscillatory behavior of bubble populations is characterized by their underlying differential equations, allowing the construction of non-oscillatory amplitude functions that can be readily integrated.
This approach requires knowledge of an appropriate bubble dynamics model, for which we employ a linearization, though the method is not entirely restricted in this sense.

The proposed method offers advantages over traditional quadrature approaches.
The computational cost remains constant as the integrands become more oscillatory with time.
Traditional methods require more quadrature points as the oscillation frequency grows (as time evolves).
The method offers a strategy for managing the complex oscillatory behavior that arises in polydisperse bubble dynamics.

The remainder of this paper is organized as follows.
\Cref{s:ensemble} presents the ensemble-averaged equations for bubbly flows and identifies the specific moments required for model closure.
\Cref{s:levin} introduces Levin's collocation method and mathematical technique.
\Cref{s:verif} verifies and demonstrates the method using a canonical oscillatory integral.
\Cref{s:bubble} develops the application of Levin's method for polydisperse bubble dynamics.
The development involves deriving the required differential equations and amplitude functions.
\Cref{s:results} presents computational results demonstrating the efficiency and accuracy of the proposed approach.
The paper concludes with a discussion of limitations and potential extensions to more complex two-phase dispersed flow problems.

\section{Ensemble averaging}\label{s:ensemble}

This section sets the mathematical framework for ensemble-averaged bubbly flow modeling and identifies the specific statistical moments that require efficient computation.
The approach follows the theoretical framework developed by \citet{zhang94} and \citet{ando10}, where the complex dynamics of individual bubbles are replaced by their statistical properties.

\subsection{Moment-based closure}

Ensemble-averaged models for bubbly flows rely on statistical moments of the bubble population to achieve closure of the governing equations.
These models replace the tracking of individual bubbles with ensemble-averaged quantities that represent the important physics of the multiphase system.

The challenge in ensemble-averaged modeling is that macroscopic flow properties depend on statistical moments of the bubble population.
For example, the mixture pressure includes contributions from bubble dynamics:
\begin{gather}
    p_{m} = p_l + \alpha \left(
		\frac{\overbar{R^3 p_{bw} }}{\overbar{R^3}} - \rho \frac{ \overbar{ R^3 \dot{R}^2 }}{ \overbar{R^3} }
	\right),
    \label{e:pressure}
\end{gather}
where $p_l$ is the liquid pressure, $\alpha$ is the void fraction, and the overbar notation denotes ensemble averaging, or moments, of the sub-grid bubbles.
The quantities $R$ and $\dot{R}$ represent the instantaneous bubble radius and radial velocity, $p_{bw}$ is the bubble wall pressure, and $\rho$ is the liquid density.

The evolution of void fraction also depends on bubble population statistics:
\begin{gather}
    \frac{D\alpha}{Dt} = 3 \alpha \frac{ \overbar{R^2 \dot{R} }}{ \overbar{R^3} },
    \label{e:alpha}
\end{gather}
where $D/Dt$ represents the material derivative following the flow~\citep{ando10}.
These expressions illustrate how macroscopic flow properties are coupled to statistical moments of the bubble population, which require moment computation.

\subsection{Statistical moments}

The computation of the moments requires integration over the bubble size distribution.
For a polydisperse bubble population with bubbles of equilibrium size $R_o$ with probability density function $f(R_o)$, the general form of these integrals is
\begin{gather}
    \overbar{R^i \dot{R}^j} = \int_0^\infty f(R_o) R^i(t,R_o) \dot{R}^j(t,R_o) \, \dd R_o.
    \label{e:general_moment}
\end{gather}

For the present analysis, we use a log-normal distribution:
\begin{gather}
    f(R_o) = \frac{1}{R_o \sigma_p \sqrt{2\pi}} \exp\left( \frac{- (\log R_o - \mu)^2}{2 \sigma_p^2} \right),
    \label{e:lognormal}
\end{gather}
where $\mu = 1$ and $\sigma_p = 0.7$ are the distribution parameters in log-space, with $\mu$ representing the mean of $\log R_o$ and $\sigma_p$ the standard deviation of $\log R_o$.
The log-normal distribution and its parameters are chosen for their physical realism and are illustrated in \cref{f:classes}~(right).
Experimental observations suggest bubble size distributions often exhibit approximately log-normal characteristics in many engineering applications~\citep{colombo21}.
In practice, the infinite integration domain is truncated to a finite interval $[R_{o,\min}, R_{o,\max}]$ where the distribution tail becomes negligible (typically where $f(R_o) < 10^{-6} \max f$, which we use herein).

\subsection{Governing equations}

To appreciate the moment-based closure challenge more fully, we consider the ensemble-averaged governing equations for bubbly flows~\citep{bryngelson19,ando11,zhang94,zhang94b}.
The macroscopic flow evolution is described by conservation of mass, momentum, and bubble number density.
The mixture continuity equation is
\begin{gather}
    \frac{\partial \rho_m}{\partial t} + \nabla \cdot (\rho_m \bu) = 0,
    \label{e:continuity}
\end{gather}
where $\rho_m = (1-\alpha)\rho_l + \alpha\rho_g$ is the mixture density, with $\rho_l$ and $\rho_g$ being the liquid and gas densities.
The mixture momentum equation is
\begin{gather}
    \frac{\partial (\rho_m \bu)}{\partial t} + \nabla \cdot (\rho_m \bu \otimes \bu) = -\nabla p_m + \nabla \cdot \boldsymbol{\tau}
    \label{e:momentum}
\end{gather}
where $p_m$ is the mixture pressure given by \cref{e:pressure} and $\boldsymbol{\tau}$ is the viscous stress tensor.

The coupling between macroscopic flow and bubble population statistics becomes explicit in the transport equations for void fraction and bubble number density.
The void fraction evolution is governed by
\begin{gather}
    \frac{\partial \alpha}{\partial t} + \nabla \cdot (\alpha \bu) = 3 \alpha \frac{\overbar{R^2 \dot{R}}}{\overbar{R^3}},
    \label{e:void_transport}
\end{gather}
which represents the evolution of the gas volume fraction due to bubble expansion and contraction.
The bubble number density satisfies
\begin{gather}
    \frac{\partial n}{\partial t} + \nabla \cdot (n \bu) = 0,
    \label{e:number_transport}
\end{gather}
where $n$ is the number density of bubbles per unit volume, assuming no coalescence or breakup processes.

This system of equations reveals the challenge addressed in this work.
The macroscopic flow evolution depends on statistical moments of the bubble population and must be computed efficiently.
The pressure and void fraction fields require moments of the bubble radii and their velocities.
Momentum exchange involves moments of bubble velocities and forces, as expected.
Computing these moments for polydisperse acoustically excited bubble populations motivates the use of an efficient computational technique.

Shown in \cref{e:pressure,e:alpha}, model closure requires the computation of four specific statistical moments of the bubble population:
\begin{gather}
    \overbar{ R^3 \dot{R}^2},  \;
    \overbar{R^3}, \;
    \overbar{R^2 \dot{R} }, \;
    \overbar{R^3  (R_o/R)^{3\gamma} },
    \label{e:rqmoments}
\end{gather}
where $R_o$ is the equilibrium bubble radius and $\gamma$ is the polytropic index (typically $1.4$).
These moments represent the statistical properties of the bubble population that influence the macroscopic flow behavior.

\subsection{Computational challenge}

\begin{figure}[htbp]
    \centering
    \includegraphics{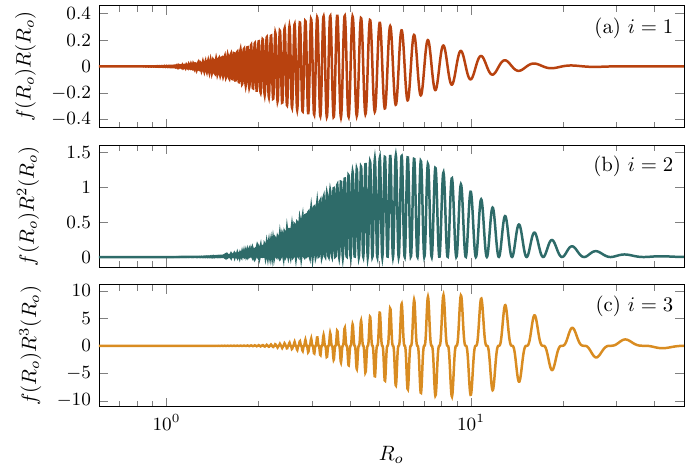}
    \caption{
        Example integrands $f(R_o) R^i(t,R_o)$ at $t=200$ for different moment orders $i$, showing the oscillatory behavior that emerges from polydisperse bubble populations.
        The complex oscillatory structure makes traditional quadrature methods computationally expensive.
    }
    \label{f:integrands}
\end{figure}

The computational challenge arises from the oscillatory nature of the bubble dynamics.
Shown in \cref{e:solution,e:velocity}, bubbles with different equilibrium radii $R_o$ oscillate at different natural frequencies $\omega_d(R_o)$.
These frequencies are typically incommensurate, and the superposition of oscillations from the polydisperse bubble population creates a likewise oscillatory structure in the integrands of \cref{e:general_moment}.
\Cref{f:integrands} illustrates this phenomenon by showing the integrands $f(R_o) R^i(t,R_o)$ at a representative time $t = 200$ for different moment orders $i$.
Herein, we focus on moments of this form, $f(R_o) R^i(t,R_o)$, which serve as a representative case of the emergent oscillatory behavior.

\section{Levin's collocation method}\label{s:levin}

\subsection{Problem statement}

We evaluate an integral of the general form
\begin{gather}
    I = \int_a^b q(x) \dd x
    \label{e:int}
\end{gather}
where the integrand $q(x)$ is highly oscillatory with respect to $x$.
Levin's method separates the oscillatory behavior from the smooth amplitude.
The approach represents the oscillatory part with a chosen basis $\bw$ that satisfies a known differential equation.
Levin showed that if $q$ decomposes appropriately, we can trade numerical integration of \cref{e:int} for the numerical solution of a differential equation~\citep{levin96}.
If the method is used appropriately, the differential equation has non-oscillatory solutions, making its solution economic compared to integrating the oscillatory integrand.

\subsection{Algorithm}

If one can decompose the integrand as 
\begin{gather}
    q(x) = \langle \bg, \bw \rangle = \sum_{i=1}^m g_i(x) w_i(x)
    \label{e:decomp}
\end{gather}
where $\langle \cdot, \cdot \rangle$ denotes the inner product over an $m$-dimensional vector space, then we can evaluate \cref{e:int} as 
\begin{gather}
    I = \int_a^b \langle \bg, \bw \rangle \, \dd x = \langle \bp, \bw \rangle \bigg|_{x=a}^b
    = \sum_{i=1}^m p_i (b) w_i(b) - \sum_{i=1}^m p_i (a) w_i (a) 
    \label{e:goal}
\end{gather}
where $m$ is the length of $\bw$, and $\bp$ are to be found.

This situation is preferable to solving \cref{e:int} via standard integration rules when the decomposition \cref{e:decomp} is chosen such that $\bw$ isolates the oscillatory part. 
Thus, $\bg$ is constructed to be non-oscillatory (called the amplitude).
The oscillatory part $\bw$ must satisfy a linear system of first-order ordinary differential equations (ODEs):
\begin{gather}
    \bw'(x) = \bA(x) \bw(x),
    \label{e:deriv}
\end{gather}
where $\bA(x)$ is a $m \times m$ non-oscillatory matrix (called the differential matrix).

We see this construction as follows.
Differentiating both sides of \cref{e:goal} trades the integral for a differential equation:
\begin{gather}
    \langle \bg, \bw \rangle = \langle \bp, \bw \rangle',
    \label{e:goal2}
\end{gather}
with derivatives on both $\bp$ and $\bw$.
Using \cref{e:deriv} we lift the derivative from $\bw$ as:
\begin{align}
    \langle \bg,\bw \rangle = \langle \bp,\bw \rangle' 
        &= \langle \bp',\bw \rangle + \langle \bp,\bw' \rangle  \nonumber\\
        &= \langle \bp',\bw \rangle + \langle \bp, \bA \bw \rangle \\
        &= \langle \bp' + \bA^\top \bp, \bw \rangle \nonumber
\end{align}
and so
\begin{gather}
    \bp' + \bA^\top \bp = \bg,
    \quad \text{or} \quad
    p_j'(x) + \sum_{i=1}^m p_i(x) A_{ij}(x) = g_j(x)
    \quad \text{for } j = 1,\dots,m.
    \label{e:alt}
\end{gather}
The solution to $\bp$ is not oscillatory in $x$ because $\bA$ and $\bg$ are not, and so comparatively cheap to evaluate.

In the use case for this work, the rapidly oscillatory content is carried by the basis built from the bubble radius and radial velocity.
The differential matrix and amplitude depend only on slowly varying mechanical coefficients, so the Levin system remains non-oscillatory even when the raw integrands (see \cref{f:integrands}) are highly oscillatory.

\subsection{Spectral collocation}

We approximate $\bp$ via $n$-point collocation with linearly independent basis functions $u_k(x)$, where $k = 1,\dots,n$.
We use $u_k(x) = T_k(x)$, where $T_k$ is the $k$-th order (scaled and shifted to the interval $[a,b]$) Chebyshev polynomial of the first kind.
Thus, we approximate $\bp$ as
\begin{gather}
    p_i(x) = \sum_{k=1}^n c_{ik} u_k(x) \quad \text{for } i = 1,\dots,m,
    \label{e:p}
\end{gather}
where $\bc$ are unknown coefficients.
Substituting \cref{e:p} into \cref{e:alt} gives
\begin{gather}
    \sum_{i=1}^m A_{ij}(x_l) \sum_{k=1}^n c_{ik} u_k(x_l) + \sum_{k=1}^n c_{jk} u_k'(x_l) 
    = g_j(x_l) 
    \quad \text{for } j=1,\dots,m, \; l=1,\dots,n,
\end{gather}
where 
\begin{gather}
    x_l = \frac{b-a}{2} \cos \left(\frac{l \pi}{n}\right) + a \quad \text{for } l=1,\dots,n
\end{gather}
are the collocation points and $u_k'(x) = T_k'(x) = k U_{k-1}(x)$ where $U_k$ is the $k$-th Chebyshev polynomial of the second kind.
The notation $n$ for the total number of collocation points is also referred to as $N$ in the figures of \cref{s:verif,s:results} for consistency with trapezoidal rule results. 
However, for computational purposes, the interpretation is the same.
The $m\,n$ linear system is the collocation conditions for $\bp$, which determine the $m\,n$ coefficients of $\bc$.
Given these, $I$ is approximated via \cref{e:goal}.

\section{Bessel oscillator example}\label{s:verif}

To verify the implementation of Levin's collocation method and its appropriate use for bubble dynamics, we apply it to a canonical oscillatory integral involving Bessel functions.
The test case also enables the evaluation of the method's accuracy and efficiency compared to traditional quadrature.

Consider the integral
\begin{gather}
    I = \int_1^2 \frac{1}{x^2 + 1} J_k(rx)\dd x
\end{gather}
for Bessel function of order $k$ with parameter $r$, which is increasingly oscillatory in $x$ for increasing $r$.
The recurrence relation for the Bessel function is
\begin{align}
    J_{k-1}' (x) &= \frac{k-1}{x} J_{k-1}(x) - J_k (x), \\
    J_k' (x)     &=  J_{k-1}(x) - \frac{k}{x}J_k (x).
\end{align}
Thus, we require $m = 2$ to determine a linear differential system with 
\begin{gather}
    \bA = \begin{bmatrix}
        (k-1)/x & -r \\
            r & -k/x 
    \end{bmatrix},
    \quad
    \bw = \begin{bmatrix}
            J_{k-1}(rx) \\
            J_k(rx) 
    \end{bmatrix},
    \quad \text{and} \quad
    \bg = \begin{bmatrix}
        0 \\
        1/(x^2 + 1) 
    \end{bmatrix},
\end{gather}
where the dimension $m = 2$ corresponds to the two-part oscillatory basis $\bw = [J_{k-1}(rx), J_k(rx)]^\top$ that satisfies the required differential equation $\bw' = \bA \bw$.

\begin{figure}[htbp]
    \centering
    \includegraphics[]{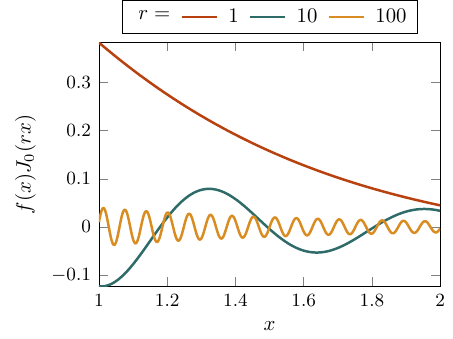}
    \caption{Integrands for the Bessel oscillator verification case showing $J_k(rx)/(x^2 + 1)$ for different oscillation parameters $r$. As $r$ increases, the integrand becomes more oscillatory, demonstrating the challenge for traditional quadrature methods.}
    \label{f:verif-int}
\end{figure}

\Cref{f:verif-int} shows the integrand $J_k(rx)/(x^2 + 1)$ for different values of $r$.
As $r$ increases, the integrand becomes more oscillatory, making traditional quadrature methods increasingly inefficient.
The Levin method maintains accuracy by separating the oscillatory Bessel function from the smooth amplitude function $1/(x^2 + 1)$.

\begin{figure}[htbp]
    \centering
    \includegraphics[]{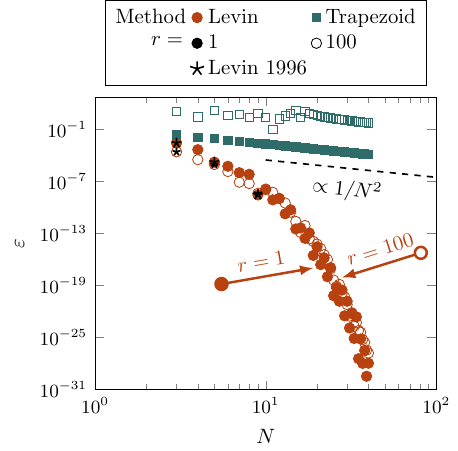}
    \caption{
        Convergence comparison between the Levin collocation method and trapezoidal rule for the Bessel oscillator integral.
        The Levin method demonstrates exponential convergence and requires orders of magnitude fewer function evaluations to achieve comparable accuracy, especially for highly oscillatory integrands (larger $r$).
    }
    \label{f:verif}
\end{figure}

\Cref{f:verif} shows the convergence behavior of the Levin collocation method and traditional trapezoidal rule quadrature.
The results are in the error $\eps$, which is the absolute difference between the exact and approximate solutions.
We demonstrate that the Levin method achieves accuracy with fewer function evaluations, particularly for highly oscillatory (large $r$) cases.
The Levin method requires fewer points than the trapezoidal rule to achieve the same accuracy for moderate oscillation frequencies.
For highly oscillatory integrands, this advantage increases to several orders of magnitude.
This comparison is made against the trapezoidal rule, which represents a basic quadrature approach.
In this case, we do not compare to Gauss--Legendre or other high-order integration schemes, as they have the same limitation, requiring an exceedingly large number of quadrature points before entering their asymptotic convergence regime.

\section{Bubble dynamics model}\label{s:bubble}

The dynamics of a spherical bubble in an infinite liquid medium are governed by the Rayleigh-Plesset equation, originally derived by \citet{rayleigh17} and later extended by \citet{plesset49}, with detailed treatment given by \citet{brennen95}:
\begin{gather}
    R \ddot{R} + \frac{3}{2} \dot{R}^2 + \frac{4}{\Rey} \frac{\dot{R}}{R} = 
    - \frac{2}{S R_o} \left( \frac{R_o}{R} - \left( \frac{R_o}{R} \right)^{3\gamma} \right) 
    + \Ca \left( \left( \frac{R_o}{R} \right)^{3\gamma}  - 1 \right)
    - C_p(t),
    \label{e:rpe}
\end{gather}
where $R(t)$ is the instantaneous bubble radius and the dot notation denotes time derivatives.
The dimensionless parameters are the Reynolds number $\Rey = \rho U R_o/\mu$, based on equilibrium radius and characteristic velocity, surface tension coefficient $S = 2\sigma/(\rho U^2 R_o)$, cavitation number $\Ca = (p_o - p_v)/(\rho U^2/2)$, polytropic index $\gamma$, and equilibrium bubble radius $R_o$.
Here $\rho$ is the liquid density, $\mu$ is the dynamic viscosity, $\sigma$ is the surface tension, $U$ is a characteristic velocity scale, and $p_v$ is the vapor pressure.
In practice, these parameters are chosen to be appropriate for the application; however, the present method is agnostic to their choice, as long as the dynamics are approximately linear and under-damped.
The dimensionless pressure that excites the bubble dynamics is
\begin{gather}
    C_p(t) = \frac{ p_\infty (t) - p_o}{p_o},
\end{gather}
where $p_\infty(t)$ is the far-field pressure and $p_o$ is the reference pressure.

The Rayleigh--Plesset equation \cref{e:rpe} can be linearized for small-amplitude oscillations about equilibrium and moderate driving pressures ($|\widetilde{R}/R_o| < 1$).
The approach is thus most appropriate for cavitation inception and moderate bubble dynamics.
The present model neglects bubble translational motion (added mass, drag forces), bubble--bubble interactions (coalescence, breakup, Bjerknes forces), and assumes spatial homogeneity of the bubble population, which are discussed in \cref{s:conclusion}.

Introducing the perturbation variable $\widetilde{R} = R - R_o$, the linearized bubble dynamics become
\begin{gather}
    \ddot{\widetilde{R}} + 2 \beta(R_o) \dot{\widetilde{R}} + \omega^2(R_o) \widetilde{R} = -\frac{C_p(t)}{R_o},
    \label{e:linearized}
\end{gather}
where the damping coefficient and natural frequency are
\begin{gather}
    \beta(R_o) = \frac{4}{\Rey R_o^2} 
	\quad \text{and} \quad 
    \omega^2(R_o) = \frac{3\gamma \, \Ca}{R_o^2} + \frac{2}{S R_o^3}(3\gamma - 1).
    \label{e:coefficients}
\end{gather}

Two trends follow directly: smaller bubbles are more strongly damped, and the restoring stiffness increases as equilibrium size decreases due to gas compressibility and surface tension.
These trends set which bubble size bands dominate the transient parts of the moments.

For step forcing with $C_p(t) = C_p \cdot H(t)$ where $H(t)$ is the Heaviside function, the solution to \cref{e:linearized} is (dropping tilde notation for convenience):
\begin{gather}
    R(t;R_o) = 
        R_o + \frac{C_p}{\omega^2(R_o)} 
        \left[ 
            1 - \exp(-\beta(R_o) t) 
            \left( 
                \cos(\omega_d(R_o) t) + \frac{\beta(R_o)}{\omega_d(R_o)} \sin(\omega_d(R_o) t) 
            \right)
        \right],
    \label{e:solution}
\end{gather}
where $\omega_d(R_o) = \sqrt{\omega^2(R_o) - \beta^2(R_o)}$ is the damped natural frequency. 
Bubbles in the viscous or sub-resonant regimes are overdamped and have different, non-oscillatory solution forms.
The corresponding radial velocity is
\begin{gather}
    \dot{R}(t;R_o) = \frac{C_p}{\omega_d(R_o)} \exp(-\beta(R_o) t) \sin(\omega_d(R_o) t).
    \label{e:velocity}
\end{gather}

Following \cref{e:rqmoments}, we evaluate the statistical moments required for ensemble-averaged model closure as integrals over the bubble size distribution as \cref{e:general_moment}.
The bubble dynamics parameters used in this work are $S = 13.9$, $\Ca = 0.977$, $\gamma = 1.4$, and $\Rey = 100$, which are representative of a standard bubbly flow configuration.
Our evaluation of the Levin method is insensitive to the specific values of the bubble dynamics parameters, so long as the dynamics are approximately linear and under-damped.

To treat boundary and initial conditions, for the linearized Rayleigh--Plesset model, initial conditions enter only through exponentially decaying transients in \cref{e:solution,e:velocity}.
At long times, the moments and their asymptotic values are independent of the initial data.
At early times, different initial conditions modify only phase and amplitude but leave the differential matrix $\bA(R_o)$ and the smooth amplitude $\bg(R_o)$ in \cref{e:alt} unchanged, so the convergence and cost of the Levin collocation are unaffected.
Alternative smooth forcings alter $\bg$ but not $\bA$, preserving the decomposition.
The present work is local to the moment evaluation, and is therefore agnostic to macroscopic flow boundary conditions.

\section{Results}\label{s:results}

\begin{figure}[htbp]
    \centering
    \includegraphics{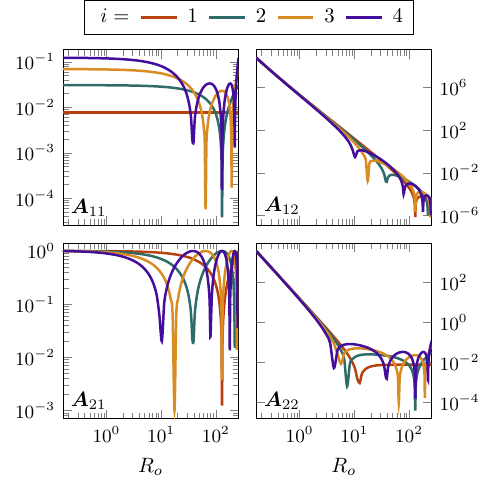}
    \caption{
        Structure of differential matrices $\bA$ for different moments $i$.
        As required for the efficient use of the Levin method, the matrices exhibit smooth, non-oscillatory behavior, with the oscillatory parts isolated into the bubble dynamics terms.
        Results correspond to the parameter values mentioned above.
    }
    \label{f:As}
\end{figure}
 
This section presents numerical results demonstrating the efficiency and accuracy of the Levin collocation method for computing statistical moments of polydisperse bubble populations.
We compare the proposed method against traditional quadrature approaches and analyze its performance across different bubble dynamics regimes.

\subsection{Differential matrix structure}

Applying the Levin method to bubble dynamics requires the construction of appropriate differential matrices $\bA$ and amplitude functions $\bg$.
\Cref{f:As} shows the structure and evolution of the four components of the $2 \times 2$ differential matrix $\bA$ for different moment orders $i = 1, \dots, 4$ in terms of the equilibrium bubble radius $R_o$.
The matrices exhibit comparatively smooth behavior across the range of equilibrium bubble sizes, making it appropriate for the Levin method.
The oscillatory or cusp behavior seen can be compared against the original integrand of \cref{f:integrands}, which is strikingly more oscillatory.
This result indicates that the oscillatory parts are primarily confined to the vector $\bw$, which contains the bubble radius and radial velocity terms, making the problem suitable for the Levin method.
Solving the corresponding system of ODEs in \cref{e:alt} is markedly less challenging for this $\bA$ than the original integrand, which is quantified next.
Consequently, the late-time growth of oscillation in the integrands does not degrade the accuracy of the Levin method.
The oscillations reside in the basis, while the differential matrix and amplitudes remain smooth in equilibrium size.

\subsection{Accuracy and efficiency comparison}

\begin{figure}[htbp]
    \centering
    \includegraphics{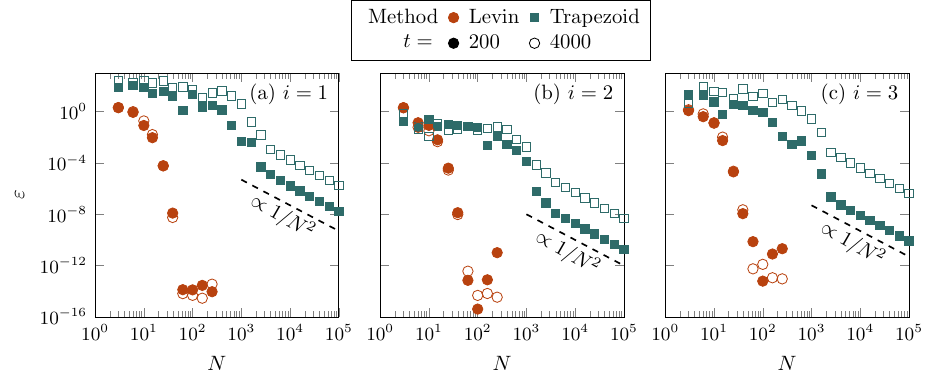}
    \caption{
        Relative error comparison between the Levin collocation method and trapezoidal rule quadrature for computing statistical moments of polydisperse bubble populations.
        The Levin method demonstrates superior efficiency, requiring 100 to 10,000 times fewer quadrature points to achieve comparable accuracy.
        Results shown for $S = 13.9$, $\Ca = 0.977$, $\gamma = 1.4$, and $\Rey = 100$.
    }
    \label{f:mom_errs}
\end{figure}

\Cref{f:mom_errs} compares the relative errors of the Levin collocation method against the traditional trapezoidal quadrature rule for the first three statistical moments ($i = 1, 2, 3$) at two different times ($t = 200$ and $t = 4000$).
The results show efficiency gains for the Levin approach across all moment orders tested.
In particular, the Levin approach achieves lower relative errors using fewer collocation points than the trapezoidal rule for all cases shown.
For the first moment $\overbar{R^3}$, the Levin method achieves a relative error of $10^{-3}$ using fewer quadrature points than the trapezoidal rule.
The advantage becomes more pronounced for higher-order moments and smaller desired errors $\eps$.
The Levin method requires fewer evaluation points for relative errors of $10^{-8}$.

The difference in performance becomes more pronounced at a later time, $t = 4000$, where the trapezoidal rule shows degraded accuracy, while the Levin method remains unaffected.
This result is a consequence of the Levin method's insensitivity to increasing oscillation frequency, provided the underlying dynamics remain of the same form, a particularly valuable advantage for bubbly flow simulations.

The convergence rate analysis shows the characteristic $1/N^2$ algebraic rate, as indicated by the dashed reference lines in all three panels.
The convergence rate is consistent with the polynomial nature of the bubble dynamics solutions.

The computational cost of the Levin method involves solving an $(m\,n) \times (m\,n)$ linear system for the polynomial coefficients at each evaluation.
Here, $m = 2$ is the dimension of the oscillatory basis (corresponding to bubble radius and velocity), and $n$ is the number of collocation points, with computational cost scaling as $O((m\,n)^3)$ for our bubble dynamics application.
The error is computed against a reference solution obtained using the Levin method with $n = 10^5$ collocation points.

While higher-order moments (e.g., $i = 4, 5, \ldots$) could be computed using the Levin method, current ensemble-averaged bubbly flow models typically require only the first few moments in the equilibrium radius dimension $R_o$.
Higher-order moments do appear in the quadrature method of moments (QMOM) formulations. However, these are primarily used for modeling coalescence and breakup processes rather than the cavitation dynamics considered here.
The present results demonstrate consistent convergence behavior across the first three moments, with no degradation observed for higher-order cases.
Given that higher-order moments share the same mathematical structure as lower-order ones (e.g., the fourth moment is also an even function), similar performance is anticipated for $i \geq 4$.
However, such moments have limited practical value in current sub-grid bubbly cavitation models, motivating our focus on the first three moments that are commonly required in engineering applications.

\subsection{Computational implications}

The efficiency gains demonstrated by the Levin method have implications for large-scale simulations of bubbly flows.
In traditional ensemble-averaged simulations, each quadrature point used for moment computation requires solving an additional advection PDE.
These requirements make the moment evaluation a computational bottleneck.
The reduction in required quadrature points translates directly to reductions in computational cost and memory requirements.

By way of example, consider a three-dimensional simulation with $N_x\,N_y\,N_z$ grid points.
In this case, for a constant error rate, the trapezoidal rule requires $N = 10^3$ quadrature points, which would entail storing and updating $10^3$-times more field variables, and so dwarfing the memory and computational requirements of the rest of the flow solver.
In contrast, the Levin method achieves equivalent accuracy with $n = 10$ collocation points, reducing this storage requirement by two orders of magnitude and making it feasible to compute the necessary moments.
As mentioned above, the Levin method's insensitivity to increasing oscillation frequency means that simulations can be extended to longer times without requiring an increasing number of quadrature points.
This property is particularly valuable for bubbly flow simulations as the bubbles are often much smaller than the carrier flow features, so they oscillate relatively rapidly, and the oscillation frequency in the integrand quickly becomes very large.

\section{Discussion}\label{s:conclusion}

\subsection{Principal conclusions}

This work applies Levin's collocation method to compute statistical moments in the dynamics of polydisperse bubble populations.
The method efficiently evaluates the highly oscillatory integrands that arise from modeling polydisperse bubble populations, a key computational challenge.

We adapted the Levin method to the context of bubble dynamics governed by linearized Rayleigh--Plesset equations.
The adaptation required the construction of differential matrices and amplitude functions that separate the oscillatory bubble dynamics from the non-oscillatory statistical distribution.
The proposed method reduces the number of required quadrature points compared to traditional methods, depending on the desired accuracy.
This reduction represents an improvement in computational efficiency for moment-based bubbly flow simulations, in some cases by several orders of magnitude.
Unlike conventional quadrature methods, the Levin approach has a constant computational cost as the bubble population displays more oscillatory behavior (at later times).
This property is useful for long-time simulations and transient bubble dynamics problems.
The cases presented represent canonical test problems for polydisperse bubble dynamics, demonstrating the method's effectiveness on the standard benchmarks used for bubbly flow modeling.

Solving the linear system of equations that the Levin method entails is about 10-times more expensive, in terms of wall time, than conducting traditional Trapezoidal rule quadrature or similar Newton--Cotes or Gauss-type quadrature rules.
However, the principal value of the work is that the Levin method foregoes the formation of PDE systems that need to be solved at each quadrature point (or Levin point).
As seen from our results, one can now conduct a simulation of the coupled Navier--Stokes equations and the Levin system (isolated to the Levin points for the moments) for the advected moments at a small number of points, or advection PDEs, when compared to traditional quadrature methods.
Solving such a large system of PDEs, even uncoupled, adds expense that readily dwarfs the cost of the Levin method itself by factors of, as aforementioned, at least two orders of magnitude for the test cases we present.

In brief, detuning-driven phase mixing, combined with viscous damping, extinguishes the oscillatory contributions to the required moments after forcing.
The Levin formulation is effective because it separates those oscillations from the smooth mechanical coefficients.
In more applied settings, we recommend pairing the present integration method with standard Euler--Euler boundary treatments, including specifying the acoustic pressure or impedance at boundaries, enforcing a no-flux condition for number density at solid walls, and advecting a measured or modeled size distribution.

\subsection{Limitations of the present work}

The method presented here requires knowledge of the underlying bubble dynamics equations, which limits its applicability to cases with analytical or semi-analytical solutions.
Constructing appropriate differential matrices becomes more complex for higher-dimensional problems or systems with multiple internal coordinates.
The linearized dynamics restrict applicability to small-amplitude oscillations (moderate cavitating flows), and the neglect of bubble--bubble interactions, coalescence/breakup, and translational motion limits the physical scope compared to full population balance approaches.

For particle-laden flows, the current work offers no advantage over traditional quadrature methods.
The Levin technique still distinguishes between oscillatory and non-oscillatory features; however, as long as no physical processes introduce oscillatory integrands, the result will match traditional quadrature methods without providing a computational advantage. 
If the size distribution function is time-dependent, then the Levin method would likely need to be coupled with a conditional quadrature moment method.
It is unclear how such coupling should be best handled and would present a new challenge in the literature.

Despite these limitations, the Levin collocation method advances computational methods for polydisperse bubble dynamics.
It eliminates a computational bottleneck in ensemble-averaged bubbly flow simulations, allowing for the efficient computation of statistical moments.
When marshaled appropriately, the method enables large-scale multiphase flow simulations that were previously computationally expensive or intractable.

\subsection{Prospects for extension}

Extending the method to nonlinear cases for bubble dynamics broadens its applicability to more extreme cavitation or dynamic conditions. 
However, the implications of this work extend beyond bubble dynamics.
The approach of computing moments from underlying dynamics contrasts with population balance methods that evolve moments through transport equations~\citep{marchisio13,fox08,bryngelson23}.
While transport-based methods avoid the oscillatory integration challenge, they require closure assumptions for unclosed moments, whereas the direct integration approach provides exact moments.
For integration into Euler--Euler solvers, the Levin method reduces the computational cost of such moment evaluation, with the trade-off between an $O((m\,n)^3)$ cost of solving a linear system against the $O(N)$ cost of traditional quadrature.

While the present work focuses on radial bubble dynamics and the efficient computation of statistical moments arising from polydisperse populations undergoing radial oscillations, the Levin method could be extended to include coalescence and breakup processes, among many others.
Coalescence and breakup processes are important phenomena in bubbly flows~\citep{zhang2024bubble,preso24}, but these effects operate on different time scales and through different physical mechanisms than the high-frequency radial dynamics considered here.
The ensemble-averaged methods used for dynamics dominated by radial bubble motion are largely uncoupled from coalescence and breakup effects.
Incorporating such phenomena into the Levin collocation framework represents a valuable extension of this work, though the interaction between these models and the Levin method has not yet been explored.
Further, even the traditional coupling of cavitation models with coalescence and breakup processes has not been well studied in the current literature.

\section*{Acknowledgments}

The author acknowledges fruitful discussion of this work with Prof.~Florian~Sch\"afer, the use of ACCESS-CI under allocation TG-PHY210084, OLCF resources under allocation CFD154, and support from the US Office of Naval Research under grant number N000-14-22-1-2519.

\section*{Data availability}

The data and code used for this study are available upon reasonable request.

\bibliographystyle{bibsty.bst}
\bibliography{main.bib}

\end{document}